# GREEN GRID: A Web-Based E-Waste Recycling Platform

***Yashodip Jagtap[1], Aaditya Bagul[2], Om Kothawal[3] Prof. Puja Patil[4]***

[1,2,3]B. Tech Student, Department of Computer Engineering, Shri Shivaji Vidhya Prasarak Sanstha's Bapusaheb Shivajirao Deore College of Engineering, Dhule (MS) Maharashtra

[4]Assistant Professor, Department of Computer Engineering, Shri Shivaji Vidhya Prasarak Sanstha's Bapusaheb Shivajirao Deore College of Engineering, Dhule (MS) Maharashtra

## ABSTRACT

Electronic waste (e-waste) is one of the fastest-growing waste streams worldwide due to rapid technological advancements and shorter device lifespans. Improper disposal releases hazardous substances that harm the environment and human health, while valuable materials such as gold, copper, and aluminum are lost if not recycled. In 2022, approximately 62 million metric tonnes of e-waste were generated globally, but only about 22% was formally recycled. India generated around 1.751 million metric tonnes in 2023-24, with only 43% processed through authorized channels. Green Grid is a full-stack web-based platform designed to simplify and encourage e-waste recycling through an E-Dumper Locator, Green Rewards System, Insights and Awareness Hub, and Eco-Marketplace. Developed using React.js, Node.js, Express.js, SQL, Google Maps API, and JWT authentication, the platform also includes scheduled pickup services and a Recycling Impact Calculator. By combining technology, education, and incentives, Green Grid promotes responsible disposal and supports the circular economy.

## 1. INTRODUCTION

Electronic devices have become essential in modern life, supporting communication, education, healthcare, entertainment, and industrial automation. As smartphones, laptops, televisions, batteries, and household appliances are replaced more frequently, the amount of discarded electronics, known as electronic waste (e-waste), continues to grow rapidly. E-waste contains hazardous substances such as lead, mercury, cadmium, and arsenic that can pollute the environment and harm human health if disposed of improperly. At the same time, it contains valuable recoverable materials including precious metals and plastics, making proper recycling both economically and environmentally beneficial. In 2022, the world generated 62 million metric tonnes of e-waste, while India, the third-largest producer, generated approximately 1.751 million metric tonnes in 2023-24.

Despite regulations such as the E-Waste Management Rules and Extended Producer Responsibility (EPR), recycling rates remain low due to limited awareness, difficulty locating certified recycling centers, lack of incentives, and unorganized refurbishment markets. Green Grid was developed to address these challenges through a comprehensive web-based platform that integrates an E-Dumper Locator, Green Rewards System, educational resources, and an Eco-Marketplace. The platform transforms e-waste recycling into a simple, transparent, and rewarding process, encouraging citizens to adopt sustainable practices while providing valuable data and insights to recyclers and policymakers.

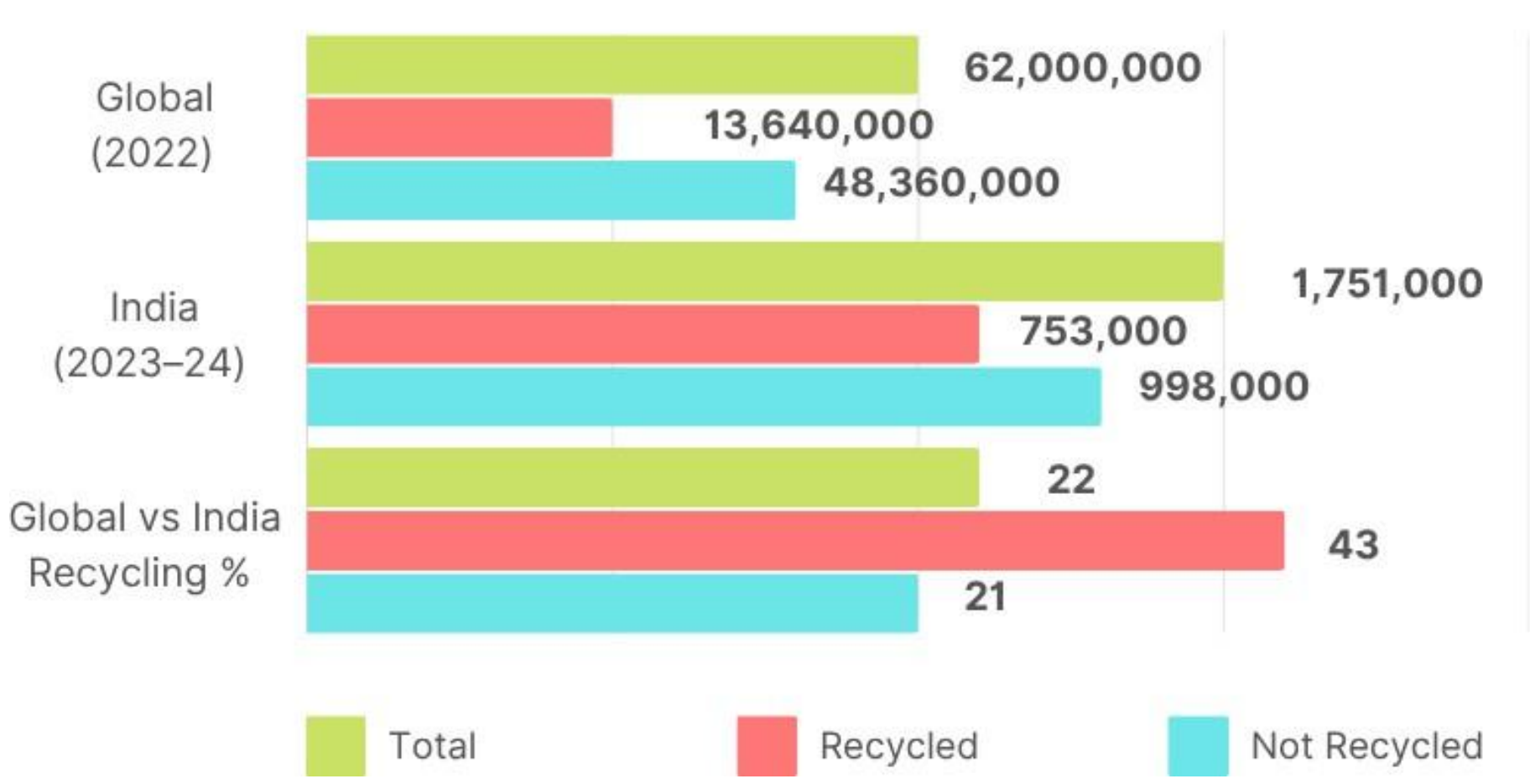

**Figure 1**: *Global versus India E-Waste Generation (2022). The chart highlights the rapid growth of e-waste in India relative to the global total.*

## 2. LITERATURE SURVEY

A variety of approaches have been proposed to improve e-waste management, including government awareness campaigns, manufacturer take-back programs under Extended Producer Responsibility (EPR), NGO-led collection drives, and private pickup services. Although these initiatives have increased public awareness, most existing systems still suffer from several limitations. Many platforms provide only static information about recycling centers, making it difficult for users to locate nearby facilities. Most systems do not offer incentives to encourage participation, educational content is often fragmented, and user engagement features such as dashboards and gamification are generally absent. In addition, recycling records are frequently maintained manually, reducing transparency, and very few systems integrate marketplaces for refurbished electronics that support circular economy practices.

Research in sustainability and environmental informatics shows that gamification, reward mechanisms, and digital engagement significantly increase citizen participation in recycling programs. Studies also emphasize the importance of extending product lifecycles through refurbishment and reuse, which reduces the demand for virgin raw materials and lowers greenhouse gas emissions. Green Grid incorporates these findings by combining location-based E-Dumper discovery, reward-based participation, awareness and educational resources, doorstep pickup scheduling, a recycling impact calculator, and an Eco-Marketplace within a single integrated platform. This comprehensive approach transforms e-waste recycling into a transparent, measurable, and rewarding process for users while generating valuable data for administrators and policymakers.

Table 1 compares Green Grid with existing e-waste management platforms such as EcoATM, Call2Recycle, and SIMS Lifecycle Services. Green Grid uniquely combines features such as E-Dumper facility discovery, gamified rewards, and environmental impact calculation, which are not offered together by competing systems. While SIMS Lifecycle Services provides doorstep pickup, it lacks citizen engagement and reward-based participation. EcoATM and Call2Recycle primarily focus on collection but do not include educational resources, impact analytics, or integrated marketplaces. By delivering all these capabilities in one unified digital ecosystem, Green Grid offers a more comprehensive and user-centric solution for modern e-waste management.

***Table 1:*** *Comparison of Green Grid vs Other Systems*

| **Parameter** | **Green Grid** | **EcoATM (USA)** | **Call2Recycle Canada/USA)** | **SIMS Lifecycle Services (Global)** |
|---|---|---|---|---|
| E-Dumpers | Yes | No | No | No |
| Doorstep Pickup | Yes | No | No | Yes |
| Gamified Rewards | Yes | No | No | No |
| AI Eco Assistant | Yes | No | No | No |
| Impact Calculator | Yes | No | No | No |

## 3. PROPOSED SYSTEM

Green Grid is proposed as an integrated full-stack web platform that formalizes and simplifies the process of e-waste recycling. The system connects citizens, recycling centers, administrators, and policymakers through a secure, scalable, and user-friendly architecture. Its primary goal is to make e-waste disposal more accessible, transparent, and rewarding while promoting environmental sustainability and circular economy practices. By digitizing each stage of the recycling process, Green Grid transforms traditional waste management into a structured and measurable ecosystem.

The platform is designed around four major objectives:

1. To help users locate nearby certified E-Dumper collection centers.
2. To encourage recycling through a Green Rewards System that awards points for responsible disposal.
3. To educate citizens about e-waste hazards, recycling methods, and sustainable practices.
4. To promote reuse and refurbishment through an integrated Eco-Marketplace.

Green Grid consists of several major modules that work together to provide a complete e-waste management solution:

1. **Authentication Module -** Handles user registration, login, password recovery, and secure session management using JWT authentication.
2. **E-Dumper Locator Module -** Displays nearby authorized recycling centers using Google Maps API and geolocation services.
3. **Green Rewards Module -** Awards Green Points for recycling activities and maintains reward and redemption history.
4. **Insights and Awareness Module -** Provides blogs, articles, and educational guides related to e-waste and sustainability.
5. **Scheduled Pickup Module -** Allows users to book doorstep pickup services if they are unable to visit a recycling center.
6. **Recycling Impact Calculator -** Estimates environmental benefits such as carbon emissions reduced, energy saved, and natural resources conserved.

The proposed system ensures that every recycling activity is digitally recorded, rewarded, and converted into measurable environmental impact. By integrating these modules into a single platform, Green Grid creates an efficient and engaging solution that benefits users, recyclers, administrators, and policymakers alike.

## 4. SYSTEM OVERVIEW AND FEATURES

Green Grid integrates multiple user-centric features that work together to create a complete e-waste management ecosystem. Each module is designed to simplify the recycling process, increase public participation, and provide measurable environmental benefits. The major features of the platform are described below.

### 4.1 E-DUMPER LOCATOR

The E-Dumper Locator helps users search for nearby authorized recycling centers using location-based services. It provides detailed information about each collection point, including the center name and address, the types of electronic materials accepted, operating hours, contact details, and the current status of the center such as whether it is open, closed, available, or full. This module removes uncertainty and makes certified recycling centers easily accessible, enabling users to dispose of their electronic waste responsibly and conveniently.

### 4.2 GREEN REWARDS SYSTEM

The Green Rewards System encourages users to recycle by awarding Green Points for environmentally responsible actions such as depositing electronic waste, scheduling doorstep pickups, and participating in awareness activities. These accumulated points can be redeemed for various rewards, including T-shirts, caps, certificates, eco-friendly merchandise, and discount coupons. This gamified approach increases user engagement and provides tangible incentives that motivate individuals to adopt sustainable recycling habits.

### 4.3 INSIGHTS AND AWARENESS HUB

The Insights and Awareness Hub serves as an educational section that informs users about the importance of proper e-waste disposal and environmentally responsible living. It contains articles explaining the hazards of electronic waste, guides on recycling best practices, sustainable lifestyle tips, and information about relevant government regulations and policies. By integrating these educational resources, the platform helps users make informed decisions and promotes long-term behavioral change toward sustainability.

### 4.4 SCHEDULED PICKUP SERVICE

The Scheduled Pickup Service provides added convenience for users who are unable to visit recycling centers in person. Through this feature, users can book doorstep collection by selecting a preferred date and time and specifying the category of electronic waste they wish to dispose of. The platform then confirms the pickup request and allows users to track its status. This feature improves accessibility and encourages participation from individuals and organizations with limited mobility or large quantities of e-waste.

### 4.5 RECYCLING IMPACT CALCULATOR

The Recycling Impact Calculator estimates the environmental benefits generated by each recycling activity. It presents measurable indicators such as carbon dioxide ($CO_2$) emissions reduced, energy saved, water conserved, and raw materials recovered. By translating recycling actions into meaningful sustainability metrics, this feature helps users understand the positive environmental impact of their contributions and reinforces the importance of responsible e-waste management.

Together, these features transform Green Grid into a practical, engaging, and scalable sustainability platform that promotes responsible e-waste disposal, public awareness, and circular economy practices.

## 5. SYSTEM ARCHITECTURE

### 5.1 SYSTEM ARCHITECTURE

Green Grid follows a three-layer architecture consisting of the Presentation Layer, Application Layer, and Data Layer. This layered design ensures modular development, scalability, maintainability, and efficient data processing. The architecture separates user interaction, business logic, and data storage, allowing each component to operate independently while communicating seamlessly. Supporting services such as Google Maps API, JWT authentication, and deployment tools further enhance the functionality, security, and reliability of the platform.

### 5.2 PRESENTATION LAYER

The Presentation Layer is responsible for the user interface and all interactions between users and the system. It is developed using React.js, HTML, CSS, and JavaScript to provide a responsive and user-friendly experience across desktop and mobile devices. This layer enables users to perform essential activities such as registering and logging into the platform, searching for nearby E-Dumper locations through map-based navigation, viewing their Green Rewards dashboard, browsing products in the Eco-Marketplace, accessing the Insights and Awareness Hub, scheduling doorstep pickups, and visualizing the environmental impact of their recycling activities.

### 5.3 APPLICATION LAYER

The Application Layer contains the core business logic of Green Grid and processes all user requests. It is developed using Node.js and Express.js, which handle routing, authentication, data validation, reward calculations, and communication with external APIs. This layer manages user authentication and authorization, executes the business rules of the system, calculates Green Reward Points, processes pickup scheduling requests, handles marketplace orders, computes environmental impact metrics, and ensures secure communication between the user interface and the database.

### 5.4 DATA LAYER

The Data Layer is responsible for storing and retrieving all application data. It is implemented using SQL and maintains structured records related to users, recycling centers, rewards, educational content, and marketplace transactions. This layer securely stores user accounts and profiles, E-Dumper location details, pickup requests, reward histories, blog posts and articles, as well as product and order information from the Eco-Marketplace.

### 5.5 SUPPORTING SERVICES

Several external services and development tools are integrated to enhance the platform's functionality and deployment. Google Maps API is used to provide geolocation and route visualization for E-Dumper discovery. JWT (JSON Web Tokens) ensures secure authentication and session management. Postman is used for API development and testing, Git and GitHub support version control and collaboration, and Render is used for deploying and hosting the application in the cloud.

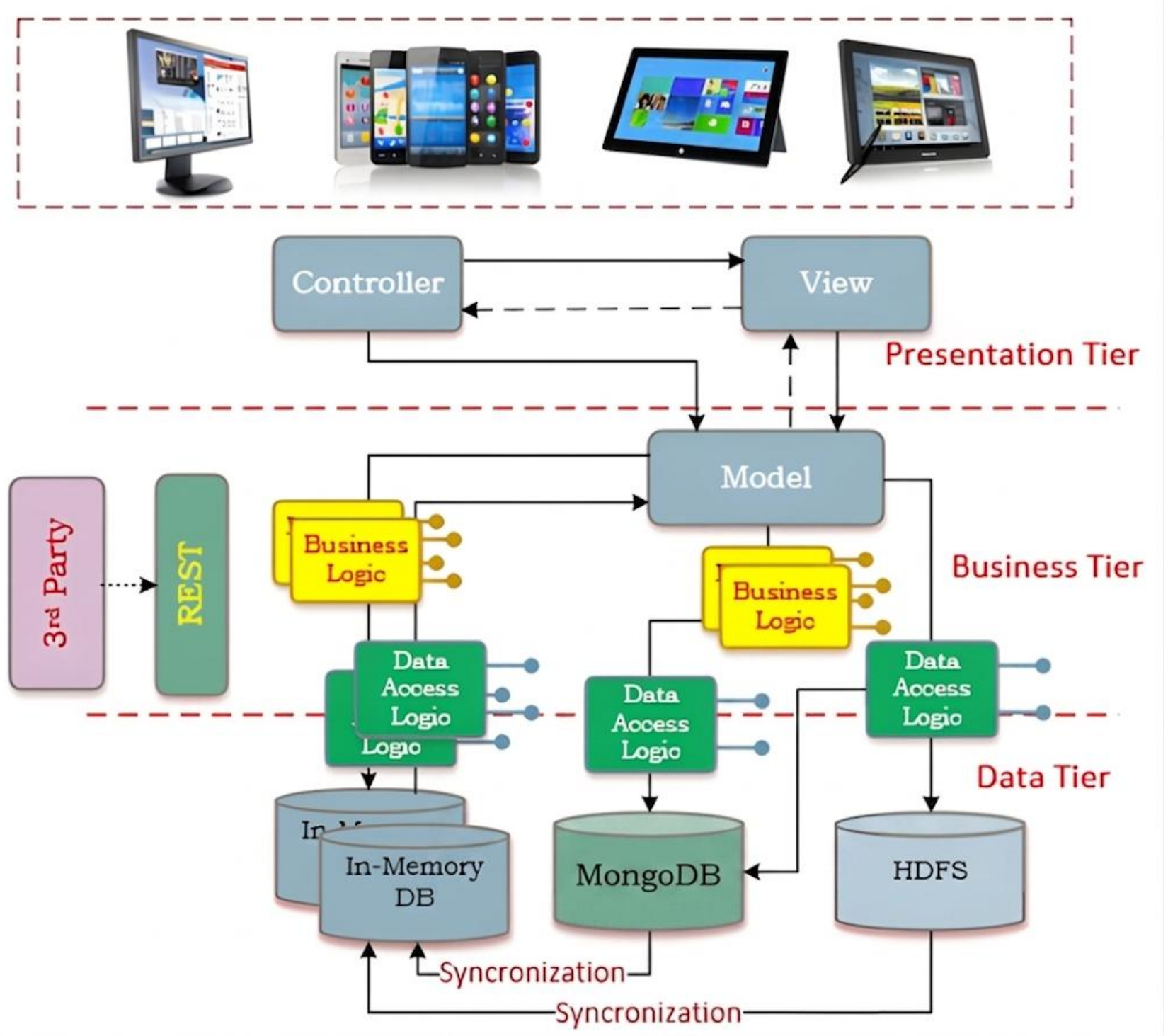

***Figure 2:*** *System Architecture of Green Grid*

## 6. WORKFLOW OF GREEN GRID

The operational workflow of Green Grid is designed to make e-waste recycling simple, accessible, and educational for users. The platform guides users through a structured process that includes locating certified recycling centers, scheduling pickups, calculating environmental impact, and accessing awareness content. Each activity is digitally recorded to ensure transparency and effective monitoring.

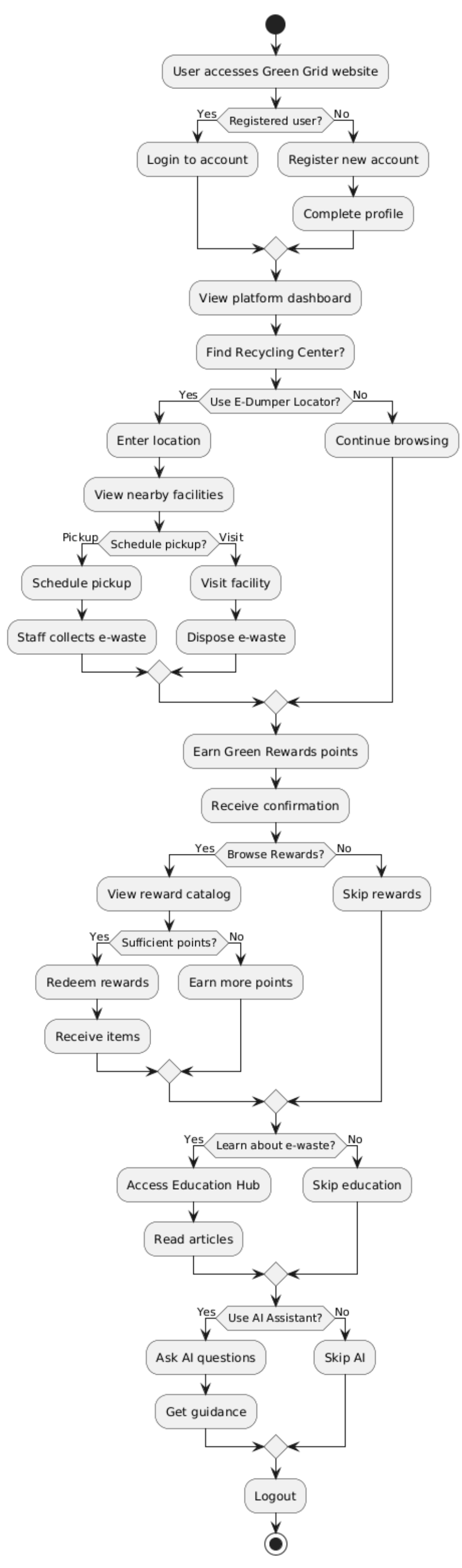

***Figure 3:*** *Workflow Diagram of Green Grid*

The workflow consists of the following steps:

1. User registers and logs into the platform using secure authentication.
2. User searches for nearby E-Dumper locations through the E-Dumper Locator integrated with Google Maps.
3. User either deposits e-waste at a certified recycling center or schedules a doorstep pickup using the Scheduled Pickup Service.
4. The system records the recycling request, pickup details, and transaction information in the database.
5. The Eco AI Assistant provides guidance on recyclable electronic items, proper disposal methods, and answers to sustainability-related questions.
6. The Recycling Impact Calculator updates environmental metrics such as $CO_2$ emissions reduced, energy saved, water conserved, and raw materials recovered.
7. Users can access educational blogs and awareness content to learn about e-waste hazards, recycling best practices, and sustainable living.
8. Administrators monitor users, E-Dumper locations, pickup requests, and environmental impact data through the administrative dashboard.

This workflow ensures that Green Grid not only enables proper e-waste disposal but also educates users, measures environmental benefits, and provides administrators with accurate tracking and reporting capabilities. By integrating six core features—E-Dumper Locator, Scheduled Pickup Service, Eco AI Assistant, Impact Calculator, Blog and Awareness, and Administrative Monitoring—Green Grid delivers a streamlined and effective solution for sustainable e-waste management.

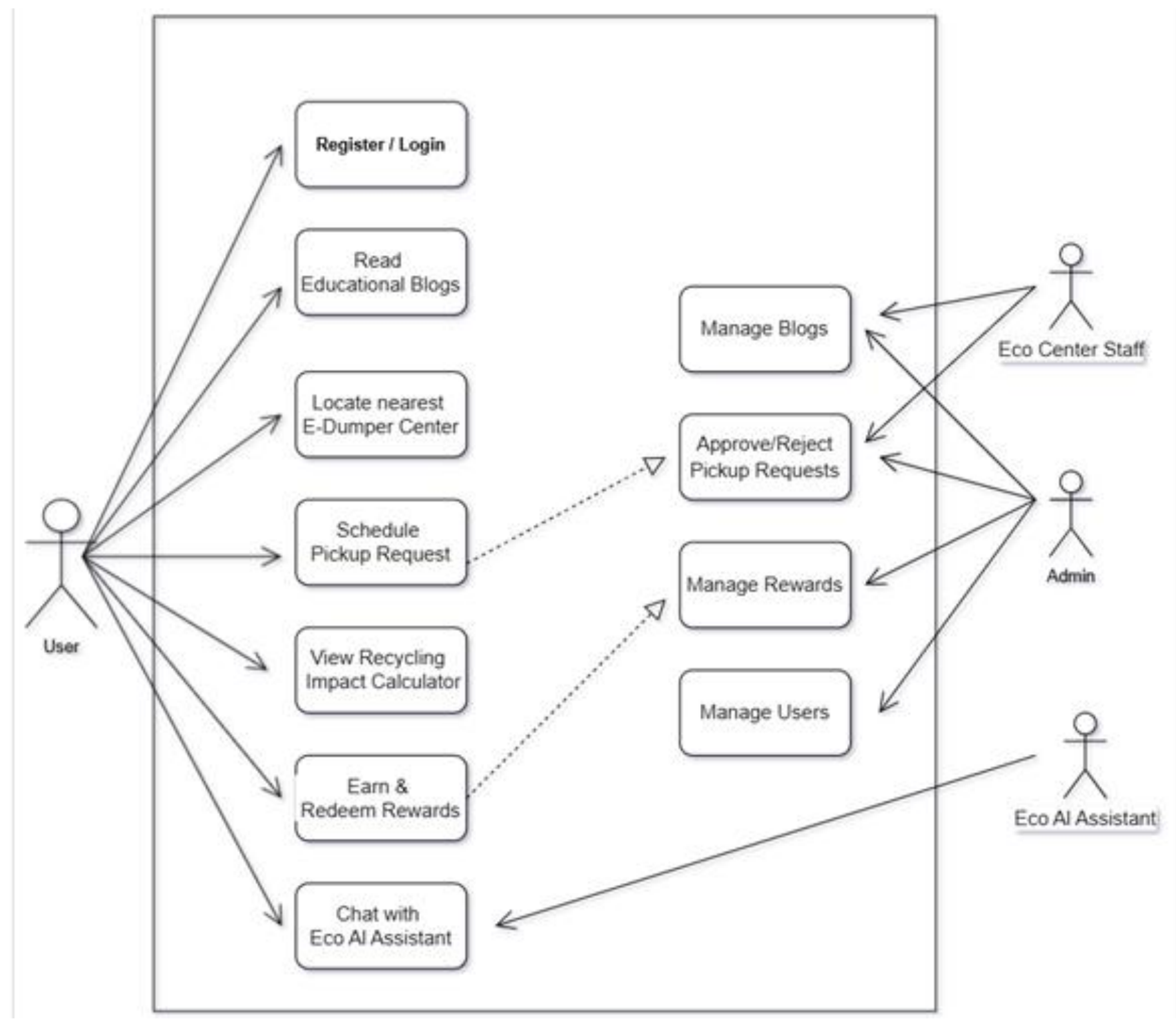

***Figure 5:** Use Case Diagram of Green Grid*

The use case diagram of Green Grid illustrates the interactions between the primary actors—User, Admin, Eco Center Staff, and Eco AI Assistant—and the core functionalities of the system. Users can register and log in, read educational blogs, locate the nearest E-Dumper center, schedule pickup requests, view the Recycling Impact Calculator, earn and redeem rewards, and interact with the Eco AI Assistant for guidance on proper e-waste disposal. Eco Center Staff and Administrators are responsible for managing blogs, approving or rejecting pickup requests, managing rewards, and maintaining user records. The Eco AI Assistant provides intelligent support by answering user queries related to recyclable items, sustainability practices, and disposal methods. This use case diagram demonstrates how Green Grid integrates user services, administrative control, and AI-based assistance into a unified platform for efficient and sustainable e-waste management.

## 7. CONCLUSION AND FUTURE SCOPE

Green Grid is a comprehensive web-based e-waste recycling platform that addresses the key challenges in modern e-waste management by integrating essential features such as the E-Dumper Locator, Scheduled Pickup Service, Eco AI Assistant, Recycling Impact Calculator, Blog and Awareness Hub, and administrative monitoring. The platform creates a seamless and engaging recycling experience by combining environmental responsibility with modern web technologies, making e-waste disposal accessible, transparent, and educational for users. Through Green Grid, citizens can locate certified recycling centers, schedule pickups, receive guidance on proper disposal, calculate the environmental impact of their recycling activities, and access informative content related to sustainability and e-waste management.

The system demonstrates how full-stack web development can be used to solve real-world environmental problems and promote circular economy principles. It not only encourages responsible behavior but also provides administrators with structured data and analytics for monitoring users, pickup requests, and environmental impact. By digitizing the entire recycling process, Green Grid transforms e-waste disposal into a measurable and efficient activity that benefits users, recycling centers, and policymakers alike.

In the future, Green Grid can be enhanced by developing native Android and iOS applications to improve accessibility, adding multilingual support for regional users, implementing QR-based deposit verification, and integrating with Extended Producer Responsibility (EPR) systems. Additional improvements may include corporate sustainability dashboards, push notifications, social leaderboards, real-time collection center capacity updates, and expansion to national and international recycling networks. With these enhancements, Green Grid has the potential to evolve into a large-scale environmental platform capable of significantly increasing recycling participation and reducing the environmental burden of electronic waste.